\documentclass[sigconf]{acmart}

\usepackage{algorithm}
\usepackage{amsmath}
\usepackage{enumitem}
\usepackage{listings}
\usepackage{xcolor}
\usepackage{multirow} 
\usepackage{microtype}

\microtypesetup{stretch=90, shrink=90} 

\AtBeginDocument{%
  }

\copyrightyear{2026}
\acmYear{2026}
\setcopyright{none}
\renewcommand\footnotetextcopyrightpermission[1]{}
\acmConference[UMAP '26]{34th ACM Conference on User Modeling, Adaptation and Personalization}{June 08--11, 2026}{Gothenburg, Sweden}
\acmBooktitle{34th ACM Conference on User Modeling, Adaptation and Personalization (UMAP '26), June 08--11, 2026, Gothenburg, Sweden}
\acmDOI{10.1145/3774935.3806176}
\acmISBN{979-8-4007-2311-7/2026/06}

\begin{document}

\title{Reflecti-Mate: A Conversational Agent for Adaptive Decision-Making Support Through System 1 and System 2 Thinking}

\author{Morita Tarvirdians}
\email{M.Tarvirdians@tudelft.nl}
\orcid{0000-0003-4246-0016}
\affiliation{%
  \institution{TU Delft}
  \city{Delft}
  \state{}
  \country{The Netherlands}
}
\author{Senthil Chandrasegaran}
\email{r.s.k.chandrasegaran@tudelft.nl}
\orcid{0000-0003-0561-2148}
\affiliation{
   \institution{TU Delft}
  \city{Delft}
  \state{}
  \country{The Netherlands}
}
\author{Hayley Hung}
\email{h.hung@tudelft.nl}
\orcid{0000-0001-9574-5395}
\affiliation{%
  \institution{TU Delft}
  \city{Delft}
  \state{}
  \country{The Netherlands}
}
\author{Catholijn M. Jonker}
\email{c.m.jonker@tudelft.nl}
\orcid{0000-0003-4780-7461}
\affiliation{%
  \institution{TU Delft/Leiden University}
  \city{Delft/Leiden}
  \state{}
  \country{The Netherlands}
}
\author{Catharine Oertel}
\email{c.r.m.m.oertel@tudelft.nl}
\orcid{0000-0002-8273-0132}
\affiliation{%
  \institution{TU Delft}
  \city{Delft}
  \state{}
  \country{The Netherlands}
}

\renewcommand{\shortauthors}{Tarvirdians et al.}

\begin{abstract}
Making high-stakes personal decisions involves cognitive, emotional, and intuitive processes, and individuals differ in how they allocate attention across these modes.
Integration of these processes has shown to benefit decision making. Yet, most current decision-support systems focus primarily on supporting cognitive aspects, rather than adapting to the individual's thinking profile to support integration of different types of thoughts.
In this study, we investigate an agent designed to encourage integration by adapting to the individual user's thought patterns. We explore its effects on participants’ perceptions of the agent and their reflective behavior, in comparison with unaided pre-reflection and a baseline agent.
In a between-subjects study (N = 128), our agent, which fostered broad and elaborated thinking, enabled more personalized reflective trajectories, elicited more integrative reflective language, and was perceived as providing stronger support for holistic reflection. In contrast, the baseline agent produced homogenized profiles dominated by cognitive language across participants.
\end{abstract}



\maketitle

\begingroup
\renewcommand\thefootnote{}
\footnotetext{Preprint. Accepted at the 34th ACM Conference on User Modeling, Adaptation and Personalization (UMAP '26), 2026.}
\addtocounter{footnote}{-1}
\endgroup

\section{Introduction}
The process of decision making is an integral part of everyday life and involves a complex interplay between cognitive, emotional, and intuitive processes~\cite{headheartgut}. Each of these processes contributes distinct and complementary information, and prior work has shown that drawing on multiple sources of judgment is associated with more effective decision making, particularly in high-stakes contexts~\cite{dotlich2006}. 
Such integration should take place during pre-decisional reflection~\cite{goodin2003when, elwyn2010deliberation, probe} phase, when people think, elaborate and organize their thoughts before committing to a decision.

\begin{figure}[t]
  \centering
  \includegraphics[width=\columnwidth]{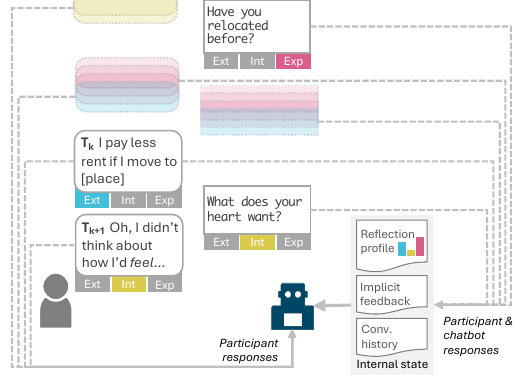}
  \caption{ReflectiMate. A possible interaction with ReflectiMate where the user begins focused on internal aspects. Guided by its internal state, ReflectiMate gently probes overlooked experiential and external aspects through exploration and exploitation questions over multiple conversational turns. Based on the current state, the agent returns to internal aspects to probe underconsidered internal reasons, illustrating its aim of supporting integrative and elaborated reflection that adapts to the user’s dynamic thought patterns and implicit feedback.}
  \label{fig:intro}
\end{figure}

However, individuals differ substantially in how they draw on and integrate these processes. Some tend to rely predominantly on a single source of judgment while neglecting others~\cite{sadler2005patterns, savioni2023}.
For example, when reflecting on a job offer that requires relocation, one person may focus heavily on anticipated excitement and fear of regret, repeatedly returning to emotional aspects while paying little attention to other considerations such as relocation logistics. 
Another person, in contrast, may engage in extensive cost-benefit analysis of salary and financial implications, while rarely considering emotional concerns such as distance from family or an intuitive sense of discomfort shaped by prior experiences.

Despite the documented benefits of such integration, people often have difficulty to do so in practice. A key challenge is limited metacognitive awareness of one’s own thinking processes~\cite{scopelliti2015, azevedo2016}, which makes it difficult to recognize how different sources of judgment influence their decision. 
For instance, people frequently overestimate the extent to which their decisions are guided by cognition while underestimating the influence of intuition and emotion~\cite{norris2011}, or fail to accurately identify which types of considerations dominate their reflection~\cite{headheartgut}. 
This lack of awareness makes it difficult for individuals to deliberately broaden or deepen their reflection in ways that integrate diverse considerations when making important decisions.

Receiving personalized support is one potential way to help individuals overcome low-awareness cognitive patterns~\cite{devine2012}. 
However, existing systems (e.g.~\cite{cognitiveAssistant, cognitiveAssistant2}) often equate reflection support with promoting cognitive reasoning. Drawing on dual-process theory~\cite{kahneman2002, evans2008dual, sloman1996dual}, emotional and intuitive considerations are frequently treated as heuristic or bias-prone (System~1) and are therefore de-emphasized or left unsupported. As a result, adaptation in such systems is typically reactive, relying only on conversational history to generate follow-up questions that steer users toward further cognitive elaboration, regardless of individual reflection patterns.

In this work, we address this gap by introducing a conversational reflection assistant that personalizes its dialog strategy to support integrative pre-decisional reflection (Fig. \ref{fig:intro}). Rather than aiming to alter users’ dominant reasoning mode ( i.e. if a person prefers cognitive reasoning we respect this rather than changing it to an emotional dominant reasoning style), our approach aims to support reflection by helping users surface potentially overlooked considerations and further elaborate underdeveloped thoughts.
We present a user model that captures how individuals distribute attention across different thought categories and how their ideas evolve over the course of a reflective interaction.
Building on this model, the agent dynamically balances exploratory questions that engage underrepresented types of thoughts with exploitative, Socratic follow-up questions~\cite{socratic, Socratic_questioning} that deepen existing ideas. 
This balance enables deliberate reflection---applying System~2-level elaboration to all thoughts, including emotional and intuitive ones typically associated with System~1---without privileging cognitive reasoning over other considerations. This implies that if a person indicates that not considering a certain aspect is deliberate, we respect the choice and do not try to 
change this.

We evaluate our approach through a user study in the context of high-stakes life decisions, examining its effects on users’ reflective language use and perceived effectiveness of the support agent.

We are answering the following research questions in this study:
\begin{itemize}
    \item \textbf{RQ1:} What differences emerge in reflective language during assisted reflection when participants interact with the experimental agent compared to a baseline agent?
    \item \textbf{RQ2:} How do participants’ perceptions of the interaction differ between the experimental agent and the baseline agent?
    \item \textbf{RQ3:} To what extent do the experimental and baseline agents preserve users’ dominant reasoning style?

\end{itemize}

\section{Related Work}
Humans are prone to a range of well-documented biases and cognitive limitations during decision making process, such as confirmation bias, framing effects, and bounded rationality~\cite{tversky1974judgment, korte_biases, bounded, boundedR1}. These limitations can distort how individuals reason and form opinions.
Therefore, a growing body of research has explored technological approaches to support people in the pre-decisional phase, when they form opinions that will guide their subsequent decision. For example, these approaches aim to help users overcome specific biases that can distort their reasoning~\cite{counteringConfirmationBias, ReducingConfirmationBias, burstFilterBubbles}.

Some studies have explored personalized interventions on information seeking behavior, as information seeking is an important aspect of the pre-decisional phase~\cite{elwyn2010deliberation}. 
These systems often model how users interact with external information to provide adaptive support that facilitates less biased behavior.
This could entail engagement with both sides (pro and con) of an argument \cite{weber2023fostering} and various topics to disrupt self-imposed filter-bubbles \cite{aicher2023SFB, aicher2022towards}.
However, these studies do not support people's own thinking process but rather focus on the process of exploring external information and opinions.

In contrast, other works have explored interventions that build directly upon individuals' own thoughts and rationales regarding their decisions. For instance, Reicherts et al.~\cite{aiHelpMe} probe users' investment rationales and, tailored to their individual portfolio profile, provide an augmented version of their thoughts that highlights overlooked facts in order to mitigate anchoring bias.
However, this work focuses on a one-shot response rather than a reflective dialog that fosters the human's own reasoning.

Beyond one-shot interactions, some approaches employ multi-turn dialog to support decision making. Hadoux et al.\cite{modelingBeliefs} model users' beliefs and concerns throughout the conversation to steer them towards particular choices. 
This works focuses therefore more on steering the user's decision rather than fostering integrative reflection.
Similarly, Chiang et al.~\cite{DevilAdvocate} work with users’ expressed thoughts by modeling the intent and stance underlying their reasoning. 
This work supports users in reflecting on the correctness of their stance, but does not aim to foster reflection that integrates cognitive, emotional, and intuitive perspectives.

Collectively, these works leave a gap: no existing system models users' thought patterns over the course of reflection to adaptively support integrative pre-decisional reasoning. 
Our work addresses this gap by introducing a conversational reflection assistant that adapts to users' evolving thought patterns to broaden and deepen reflection, while respecting 
individual reasoning preferences.

\section{Methodology}

This section presents our approach and the user study conducted to evaluate it.

\subsubsection{Reflection Phases}

In this study, the decision maker undergoes a two-phase reflection process without making a decision in either phase, focusing instead on externalizing thoughts related to the decision.
\begin{enumerate}
    \item Unaided Reflection Phase: 
The decision-maker first reflects on the decision topic independently to establish the first baseline. They proceed to the next phase once they feel they have shared sufficiently. 

    \item Assisted Reflection Phase: 
In the second phase, the decision maker interacts with a conversational agent that acts as a reflection assistant, posing questions grounded in the user’s prior unaided reflection. The interaction is process-oriented~\cite{augmenting1}, with the agent refraining from offering its own insights or solutions~\cite{thinkingAssistant}.

\end{enumerate}

\subsection{User Profile}

\subsubsection{Reflection Profile}
Once the decision maker enters the “assisted reflection phase”, our agent first constructs a computational representation of the user’s previously shared reflection (in the “unaided reflection phase”). This representation serves as the structured knowledge base guiding the agent’s subsequent decisions.

Each reflection is represented as a collection of individual thoughts, where each thought is characterized by a category, indicating the type of consideration it expresses, and by its elaborations, which serve as an indicator of its reasoning depth.
An elaboration is a justification or supporting detail that deepens a main thought rather than introducing a new one.

Formally, a reflection is represented as:
\begin{equation}
R = (T, G)
\end{equation}
where
\begin{itemize}
    \item $T = \{t_1, t_2, \dots, t_n\}$ is the set of main thoughts,
    \item $G = \{(t_i, E_i) \mid t_i \in T\}$ is the set of parent–child relations, where each $E_i$ is a set of elaborations associated with $t_i$ (a main thought).
\end{itemize}

Each main thought $t_i$ is represented as a tuple:
\begin{equation}
t_i = (\text{text}_i, \text{category}_i)
\end{equation}
where $\text{text}_i$ is the content of the main thought and $\text{category}_i \in C$ is the category of the thought, with
\begin{equation}
C = \{\text{internal}, \text{external}, \text{experiential}, \text{other}\}.
\end{equation}

We generate this representation using a locally hosted Phi-4 14B~\cite{abdin2024phi4} model (Q4\_K\_M quantization, temperature = 0), chosen after pilot testing several models that could be deployed on local servers. 

\subsubsection{Thought categories}
The category of a thought reflects the type of consideration expressed in its content. 
These categories operationalize the head–heart–gut framework~\cite{headheartgut, dotlich2006}, in combination with findings from ~\citet{klein}.
Thoughts labeled as “internal” express internally oriented considerations such as feelings or personal values. “External” thoughts refer to considerations shaped by external factors, such as situational constraints, other stakeholders, or factual information. “Experiential” thoughts draw on past experiences or prior outcomes, which have been shown to play a central role in human decision making~\cite{klein}. The “other” category captures thoughts that do not clearly fall into the above categories.

We deliberately make no claims about underlying cognitive processes, focusing instead on their behavioral correlates as expressed in language.
While these correlates relate to different sources of judgment, the relationship is not one-to-one. Internal considerations, for example, can emerge from heart or gut (emotional or intuitive processes); external considerations from head (cognitive process); and experiential considerations from gut or head (intuitive or cognitive processes).

\subsection{Adaptation Algorithm}
Once the agent has the user reflection profile, it aims to identify patterns in it. In particular, it attempts to detect: first, fixation on specific categories of thought, which can cause the user's decision to be based on narrow thinking; and second, surface-level reasoning, which can cause the user's decision to be based on fast, insufficiently considered thoughts.

To this end, the agent constructs a user's reflective pattern model based on the profile. For such modeling, two dimensions are defined:
\begin{enumerate}
    \item \textbf{Breadth} — reflecting how many distinct main thoughts the user expresses within each thought category.
    \item \textbf{Depth} — reflecting how elaborated those thoughts are through associated explanations or expansions.
\end{enumerate}

The depth of a main thought $t_i$ is defined based on its elaborative richness:
\begin{equation}
\label{eq:depth}
D_i = 1 + |E_i|
\end{equation}
where $|E_i|$ is the number of elaborations associated with $t_i$. The constant term $1$ ensures that a thought without any elaborations is assigned a depth of 1, representing the thought itself.

The breadth of reflection within a thought category $k$ is defined as:
\begin{equation}
B_k = |\{t_i \in T \mid C(t_i) = k\}|
\end{equation}
which represents the number of distinct thoughts the user produced from that perspective.

Using these two measures, we derive a cumulative indicator of the user’s fixation on a thought category:
\begin{equation}
\label{eq:fixation}
S_k = \sum_{t_i : C(t_i) = k} D_i = \sum_{t_i : C(t_i) = k} (1 + |E_i|) = B_k + \sum_{t_i : C(t_i) = k} |E_i|
\end{equation}

It increases both with the number of thoughts in a thought category (breadth) and with the number of elaborations attached to them (depth), thus reflecting the overall focus of reflection within that category.

In order to model the user's reasoning depth patterns, the following indicator is defined:
\begin{equation}
\label{eq:shallow}
\bar{D}_k =
\begin{cases}
\frac{S_k}{B_k}, & \text{if } B_k > 0 \\
0, & \text{otherwise}
\end{cases}
\end{equation}

This indicator provides an average measure of elaboration within each thought category, independent of the number of thoughts expressed, thus solely reflecting the average degree of reasoning depth within each thought category.

The reflection profile and pattern model (the indicators) are initially constructed at the start of the “assisted reflection phase” and dynamically updated after each turn to guide the agent based on the user’s current reflective state.

The agent has two main objectives in this interaction:
\begin{enumerate}
    \item To disrupt fixation patterns, thereby broadening the user’s reflection.
    \item To disrupt shallow reasoning patterns, thereby deepening the user’s reflection.
\end{enumerate}

\subsubsection{Disrupt disadvantageous thought patterns}
(1) To disrupt fixation: The agent uses the fixation indicator (Equation~\ref{eq:fixation}) to identify the categories receiving the most attention. Rather than reinforcing this focus, it redirects the user to the least explored category-analogous to uncertainty-based sampling in adaptive learning~\cite{lewis1994sequential,settles2009active}-on the assumption that these categories offer the greatest opportunity to broaden reflection.

(2) To disrupt shallow reasoning: The agent uses depth indicators (Equation~\ref{eq:shallow}) to identify underdeveloped thoughts. Categories with lower depth scores contain less elaborated reflections, and within them, thoughts with the lowest depth scores (Equation~\ref{eq:depth}) are prioritized for follow-up, maximizing potential gains in reflection depth.


\subsubsection{Reflection Enhancing Strategies}
To satisfy both objectives throughout the interaction, the agent formulates its decision process as an \emph{exploration-exploitation tradeoff}~\cite{sutton2018reinforcement}. At each turn \(t\), it decides whether to elicit new thoughts to broaden reflection (\emph{exploration}) or deepen existing thoughts (\emph{exploitation}).

This decision is guided by a continuously updated internal state \(s_t\), which includes the structured reflection profile, the user’s reflective pattern model, the history of asked questions, and the dialog context. Several signals derived from \(s_t\) inform action selection: the agent favors novel prompts to reduce repetition, prioritizes underdeveloped thoughts for exploitation, and adjusts probing based on implicit user feedback, reducing the utility of thoughts that previously elicited minimal elaboration, signaling low chance of further development.

The agent estimates the relative utility of exploration and exploitation and applies an \(\varepsilon\)-greedy policy~\cite{sutton2018reinforcement,watkins1992q}, selecting exploration with probability \(\varepsilon\) and exploitation with probability \(1-\varepsilon\):
\begin{equation}
a_t =
\begin{cases} 
\text{explore}, & \text{with probability } \varepsilon \\
\text{exploit}, & \text{with probability } 1 - \varepsilon
\end{cases}
\label{eq:epsilon_greedy}
\end{equation}

During exploration, the agent draws questions targeting underrepresented categories from a curated set of category-specific prompts (Appendix~B, peer-reviewed by three researchers for clarity), to eliminate phrasing as a potential confounding variable.
During exploitation, it generates Socratic follow-ups~\cite{socratic,Socratic_questioning} using a locally hosted Phi-4 14B model (Q4\_K\_M quantization, temperature = 0; Appendix-C), fostering elaboration without introducing solutions.

After each response, the agent updates the reflection profile: exploration responses are first categorized and added as new thought nodes, and exploitation responses are appended as child elaborations. The adaptation algorithm is then updated accordingly, ensuring subsequent decisions reflect the current reflective state.

The agent maintains conversational coherence by leveraging dialog history and favoring questions within the same category across adjacent turns, minimizing unnecessary topic shifts and supporting a natural interaction.

\subsection{User Study}

\begin{figure*}
\centering
\includegraphics[width=\textwidth]{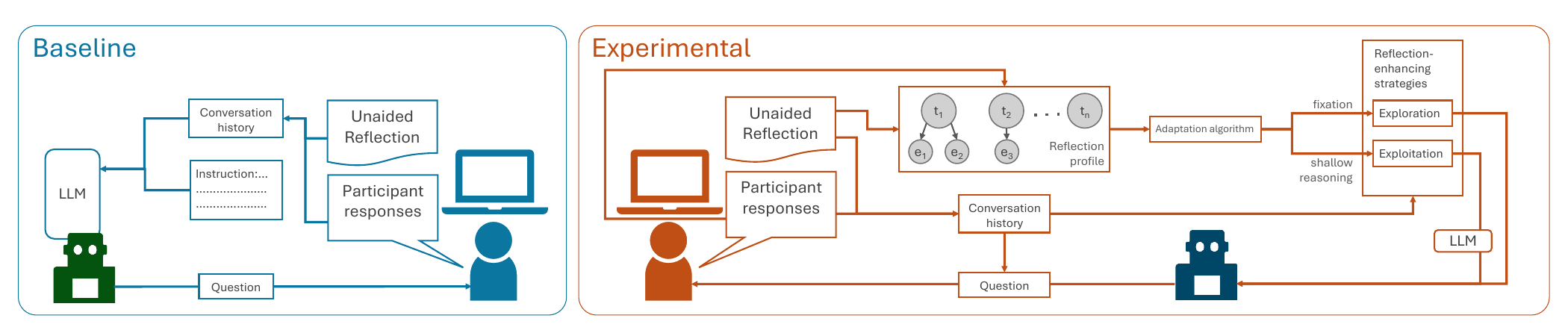}
\caption{Overview of the Conditions. 
The baseline agent selects the next prompt based on the conversational history—including unaided user reflection—and a fixed instruction specifying objectives and constraints, which are passed to the LLM for prompt generation. The experimental agent additionally uses an adaptation algorithm informed by the user’s reflection profile to identify thought patterns, and pose either an exploration or exploitation question accordingly.}
\label{fig:StudyOverview}
\end{figure*}

We conducted a between-subjects study comparing two versions of a conversational reflection assistant during the “assisted reflection phase.” (See Figure~\ref{fig:StudyOverview})

In the experimental condition, the agent’s behavior was guided by the reflection profile and adaptive strategies described above, which tracked users’ reflective patterns to decide when to broaden reflection or prompt deeper elaboration. 

The baseline agent was instructed to pursue the same high-level objectives without maintaining an explicit profile of the user’s reflection (Appendix-D).
Both agents used the same underlying language model (Phi-4 14B) and shared interaction goals to isolate the effect of profile-driven adaptive dialog management while controlling for language generation, task, and interaction context.

\subsubsection{Procedure}
The study was conducted on a web-based platform. After providing informed consent describing the study purpose, data use, and participant rights, participants completed a pre-questionnaire collecting demographic information, self-reported decision clarity (5-point Likert scale), and the short form of the Self-Reflection and Insight Scale (SRIS)~\cite{SRIS-short}.  
Participants then selected a decision topic and entered the “unaided reflection phase”, freely articulating their thoughts in text without external intervention. Once they indicated they had no further thoughts, they proceeded to the “assisted reflection phase”. Here, participants interacted with a conversational agent and were randomly assigned to the experimental or baseline condition, without being informed of the assignment. Interaction continued until participants chose to end the session, with a minimum of ten conversational turns to ensure sufficient engagement.  
After the interaction, participants rated post-statements assessing the agent’s perceived effectiveness. All statements were reviewed by three researchers for ambiguity, bias, and ethical considerations prior to the study, in line with established content validity and scale development procedures~\cite{boateng2018best, lynn1986determination}.

\subsubsection{Participants}
We recruited 128 participants (64 per condition) via the crowd-sourcing platform Prolific\footnote{\url{https://www.prolific.com}}, with sample size determined a priori using G*Power~\cite{G*,G*1} to achieve 80\% statistical power. Inclusion criteria were fluency in English and currently being in the process of making one of the big life decisions~\cite{camilleri2023investigation} listed in Appendix-A. These decisions were chosen because they represent moments when reflection is particularly crucial~\cite{mols2016informing}, given their potential long-term impact on people’s lives.
A chi-square test revealed no significant association between topic and condition ($\chi^2(13)=21.73$, $p=.060$), suggesting balanced topic distribution across conditions.

Participants were diverse in age (mean = 41.2, SD = 13.3, median = 40.1, range 18--65+), gender (66 female, 62 male), education (ranging from no diploma to doctorate), self-reflection tendencies measured with the short form of Self-Reflection and Insight Scale~\cite{SRIS-short} (mean = 46.08, SD = 7.26, median = 46), and decision clarity, self-reported on a 5-point Likert scale (median = 4).

\subsubsection{Measures}
The effectiveness of the interaction was evaluated across two dimensions:

\textit{Reflective language.} We analyzed participants’ reflections using linguistic markers associated with cognitive, emotional, and intuitive processing, quantified with the Linguistic Inquiry and Word Count dictionary (LIWC-22)~\cite{LIWC22} and operationalized with established category groupings~\cite{LIWC-cognitive, LIWC-affective-cognitive, LIWC-intuitive-affective}.
Cognitive processing included insight, causation, discrepancy, and tentative categories; emotional processing included positive and negative emotion; intuitive processing included past-tense verbs and perceptual processes (seeing, hearing, feeling).

\textit{Perceived support.} We evaluated participants’ perceptions of the agent’s effectiveness in integrating head, heart, and gut perspectives during reflection and in deepening their reflections (see Table~\ref{tab:post_statements}). Ratings were collected using 5-point Likert scales, and participants provided open-ended responses to elaborate on their evaluations.

\subsubsection{Ethical Considerations}
We obtained ethical approval for running this study from the university’s Human Research Ethics Committee (HREC). All participants provided informed consent and were compensated in accordance with the crowd-sourcing platform’s policies.
Given the sensitivity of personal decision making, the AI agent was designed to avoid directive advice and instead support reflective inquiry. When participants requested opinions or recommendations, the agent clearly stated that it could not provide advice, and participants were not asked to reach or justify a final decision.
All LLMs were deployed locally to ensure that reflection data remained on secure servers and were not shared with third parties. Data were anonymized, securely stored, and personally identifiable information was separated from reflection content; participants were informed of data handling and retention practices during consent.

\section{Results}
\subsection{Reflection Behavior}
To examine whether dialog management strategy (baseline vs experimental) influenced how users expressed and developed their reflections, we analyzed linguistic markers in users’ conversational turns during the assisted reflection phase. These markers were aggregated into three composite dimensions, cognitive, emotional, and intuitive using LIWC (see User Study–Measures).

A multivariate analysis of variance (MANOVA) conducted on users' conversational reflections revealed a significant effect of condition on the joint distribution of reflective language dimensions, Wilks' $\Lambda = 0.878$, $F(3,124) = 5.74$, $p = .001$.
\footnote{Assumption checks indicated violations of covariance homogeneity (Box's M, $p = .022$) and multivariate normality (Henze--Zirkler). With equal sample sizes, MANOVA is robust to such violations~\cite{tabachnick2019}. All four test statistics yielded consistent results ($p = .001$), including Pillai's trace, and no multicollinearity was detected (all $r < .20$).}
This result indicates that users interacting with the experimental and baseline agents differed in how cognitive, emotional, and intuitive linguistic markers were jointly expressed during assisted reflection.

\subsubsection{Distribution and Differences in Dimensions}
To characterize the nature of this effect, we examined within-condition differences among the three dimensions and between-condition differences during assisted reflection.

\begin{figure}[h]
    \centering
    \includegraphics[width=\columnwidth]{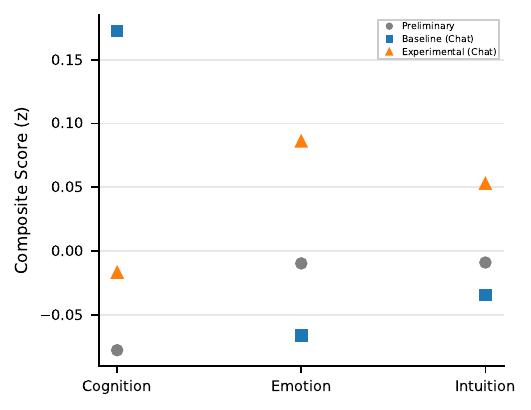}
    \caption{Transformation of Linguistic Markers by Condition. The figure illustrates mean composite z-scores for cognitive, emotional, and intuitive dimensions across Preliminary (unaided) reflections and and both conditions: Baseline and Experimental. While the Baseline condition results in a pronounced dominance of cognitive markers at the expense of other dimensions , the experimental agent facilitates multi-modal participation.}
    \label{fig:Liwc}
\end{figure}

In the interaction with the baseline agent, users’ conversational reflections exhibited significant differences across dimensions, $F = 7.53$, $p < .001$. This pattern was characterized by a pronounced dominance of cognitive linguistic markers relative to emotional and intuitive markers (see Figure~\ref{fig:Liwc}).
In contrast, no significant differences across dimensions were observed in the experimental condition, $F = 1.12$, $p = .33$, with users expressing cognitive, emotional, and intuitive markers at more comparable levels during the assisted reflection phase (see Figure~\ref{fig:Liwc}).

To quantify between-condition differences in reflective expression, we computed Cohen’s $d$ for each composite dimension. Relative to the baseline condition, users interacting with the experimental agent exhibited lower cognitive composite scores ($d = -0.51$) and higher emotional ($d = 0.35$) and intuitive ($d = 0.24$) composite scores.

These differences reflect shifts in relative salience rather than suppression of cognitive reflection. Cognitive markers remained robustly present in the experimental condition; however, emotional and intuitive markers were more strongly expressed alongside cognitive markers.

\subsection{Perceived Support}
\begin{figure}[h]
    \centering
    \includegraphics[width=\columnwidth]{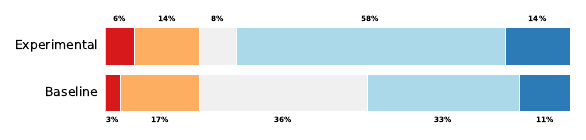}
    \includegraphics[width=\columnwidth]{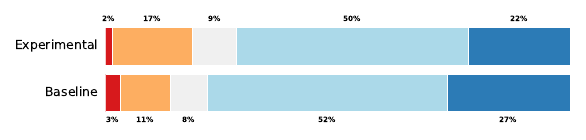} \\
    \includegraphics[width=0.8\linewidth]{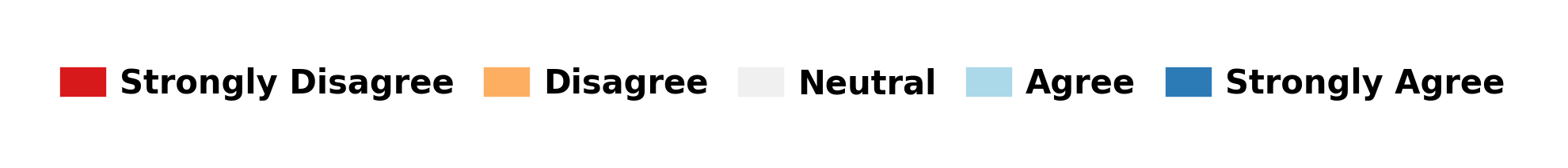}
    \caption{Distribution of participant responses for the two perceived support statements (see Appendix-E). 
    \textit{Top}: holistic integration. 
    \textit{Bottom}: elaboration depth.}
    \label{fig:likert_results}
\end{figure}

Participants rated perceived holistic integration.
As shown in Fig.~\ref{fig:likert_results}, ratings were significantly higher in the experimental condition: 72\% agreed or strongly agreed, compared to 44\% in the baseline condition. 
A chi-square test on collapsed categories (Disagree/Neutral/Agree) confirmed a significant association between condition and response, $\chi^2(2, N = 128) = 15.95, p = .002$ (Holm-corrected), with a moderate effect size ($V = .35$). 
Baseline participants also exhibited greater uncertainty, with more neutral responses (36\% vs.\ 8\%).

In contrast, ratings for thought elaboration did not differ between conditions, with high agreement in both groups (experimental: 72\%, baseline: 79\%), $\chi^2(2, N = 128) = 0.69, p = 1.0$, $V = .07$.

\subsection{Heterogeneous vs. Homogeneous Effects}
The experimental agent adapted its questioning strategy online based on the evolving reflection profile of its users. To explore whether this adaptive strategy produced heterogeneous downstream effects on reflective language use across participants, we conducted a post-hoc cluster analysis. Users were clustered based on their pre-interaction (unaided reflection phase) linguistic profiles using k-means, yielding three interpretable archetypes:

\begin{itemize}
    \item \textbf{Cluster 0 -- Reserved ($n$ = 73):} Low expression across all modes (Cog: $-0.11$, Emo: $-0.43$, Int: $-0.21$).
    \item \textbf{Cluster 1 -- Intuition-dominant ($n$ = 24):} High intuitive markers (Int: $0.91$).
    \item \textbf{Cluster 2 -- Emotion-dominant ($n$ = 31):} High emotional markers (Emo: $1.25$).
\end{itemize}

These clusters were used to investigate whether the effect of the dialogue strategy varied as a function of users' initial reflective language patterns, addressing \textbf{RQ3}.

\subsubsection{Condition Effects Varied by Initial Profile}
\begin{figure}[ht]
    \centering
    \includegraphics[width=\columnwidth]{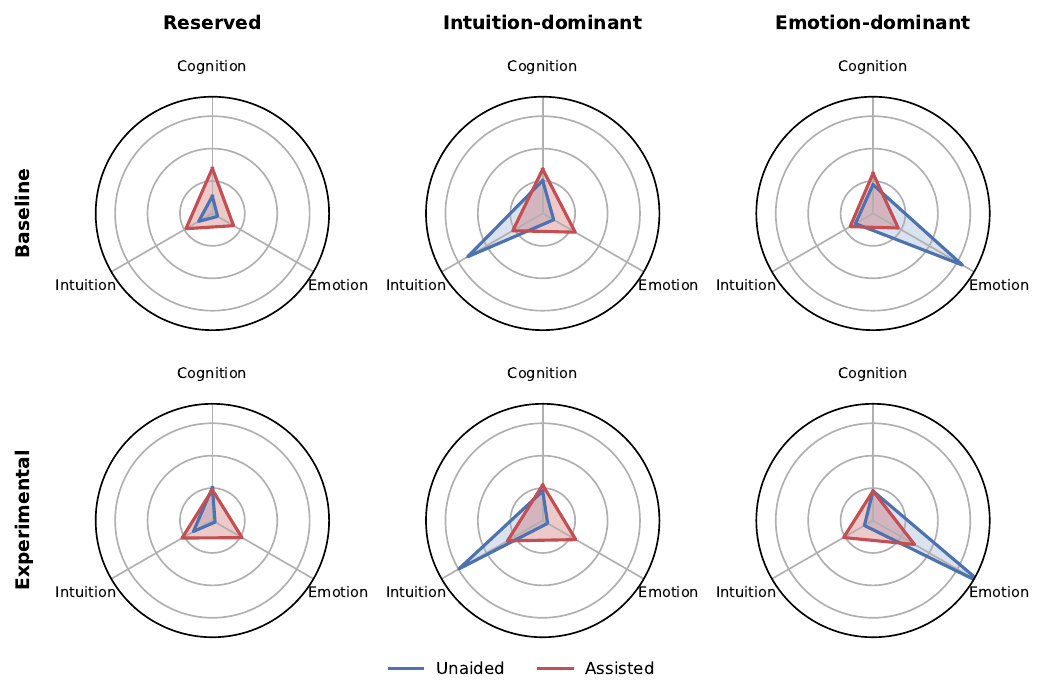}
    \caption{Reflection trajectories across clusters and conditions. Radar plots visualize the transition from unaided (blue) to assisted (red) reflection across three user archetypes. Axis scales represent standardized reflection scores (z-scores), where distance from the center indicates the relative intensity and reliance on a specific reflective language mode}
    \label{fig:radar_clusters}
\end{figure}

Figure~\ref{fig:radar_clusters} reveals that the baseline and experimental agents produced different effects depending on users' initial reflective profiles. 

Across all three clusters, the baseline agent amplified cognitive markers while reducing the relative prominence of users' initially dominant modes. In Cluster 0 (Reserved), all dimensions increased but cognition dominated ($\Delta_{\text{Cog}} = 0.43$, $d = 0.74$, $p < .001$). In Cluster 1 (Intuition-dominant), the initially prominent intuition collapsed ($\Delta_{\text{Int}} = -0.80$, $d = -1.60$, $p < .001$) as cognition and emotion increased. In Cluster 2 (Emotion-dominant), emotional expression was substantially reduced ($\Delta_{\text{Emo}} = -1.14$, $d = -1.14$, $p < .001$). Despite starting from markedly different baseline profiles, all three clusters moved toward a similar post-interaction configuration characterized by elevated cognition, indicating a homogenizing effect independent of initial reflective patterns.

In contrast, the experimental agent selectively expanded underrepresented modes while the initially dominant mode typically remained the most prominent.
In Cluster 0, emotional ($d = 0.90$, $p < .001$) and intuitive markers ($d = 0.49$, $p = .005$) increased. In Cluster 1, intuition remained the dominant mode ($d = -1.86$, $p < .001$) while emotion was scaffolded ($d = 0.90$, $p = .010$). In Cluster 2, emotional markers remained dominant ($d = -1.14$, $p < .001$) while intuition was scaffolded ($d = 0.67$, $p = .026$). Rather than converging toward a single profile, users in the experimental condition retained cluster-specific structure while exhibiting broader multi-modal participation.

\subsubsection{Differential Impact on Cognitive Markers}
The most striking difference between conditions emerged in cognitive language use. The baseline agent produced uniformly large cognitive increases across all clusters ($\Delta_{\text{Cog}}$ range: $0.17$--$0.43$, $d = 0.27$--$0.74$, all $p < .05$). In contrast, the experimental agent showed non-significant cognitive changes (all $|\Delta_{\text{Cog}}| < 0.10$, all $p > .31$), suggesting that adaptive scaffolding focused on expanding all modes rather than uniformly amplifying cognition.

\section{Discussion}
We investigated whether an adaptive conversational agent could support more integrative pre-decisional reflection than instruction-based approaches. In this section, we interpret our findings and consider their implications for designing decision support agents.

In \textbf{RQ1}, we investigated differences in reflective language use between our experimental and baseline agent. While prior work \cite{headheartgut, somatic} has argued for an integrative approach to decision making, and \cite{dotlich2006} demonstrated the benefits of such integration in real-world leadership and business scenarios, our agent was able to enhance these patterns compared to the baseline.
Although assessing task outcomes would require longitudinal studies, which are beyond the scope of the current work, our findings indicate the potential of conversational agents to support longer-term decision making.

The baseline agent's failure illuminates why computational user modeling matters. Despite identical goals (support breadth and depth across various thoughts) and access to the same conversational context, it could not achieve integration because it lacked the structured representation needed to detect when reflection was imbalanced or shallow. Instructions alone cannot substitute for this computational awareness. This suggests that effective reflection support requires not just good prompts, but principled adaptation grounded in real-time user modeling.

In \textbf{RQ2}, we explored the participants’ perceptions of the interaction, and its
results further suggest that participants not only found the agent helpful in supporting integration but also in elaborating on their thoughts.
While prior work (e.g., \cite{aicher2023SFB}) shows increased elaborateness when participants are presented with diverse arguments, our results indicate that depth of reasoning can also be increased by engaging with the underlying driving forces of participants’ own thoughts.

In \textbf{RQ3}, we investigated whether our agent could respect users dominant reasoning style while supporting integration and elaboration at the same time. We found that it could, as it produced divergent effects-maintaining the prominence of each user’s dominant mode while selectively expanding underrepresented dimensions. Importantly, these differentiated outcomes emerged even though the experimental agent was not explicitly conditioned on cluster membership, suggesting that real-time adaptive scaffolding can support personalized trajectories without requiring predefined user clusters.

This finding is particularly encouraging given prior work \cite{copilot}, which suggests that although psychological support extends beyond purely cognitive nudging toward more holistic approaches, LLM-based agents often fall short in preserving individual reasoning styles, focusing narrowly on users’ initial thoughts and asking follow-up questions that deepen but do not broaden reflection. 
Our baseline agent exemplified this limitation. Its dominance of cognitive reflective markers indicates that it primarily probed users to further rationalize their thoughts, echoing prior work equating reflection support with promoting cognitive reasoning alone~\cite{cognitiveAssistant2}. This imposed a one-size-fits-all reasoning style, driving all users toward cognitive dominance regardless of their initial reflective tendencies.
This demonstrates the potential of our approach in decision making support, where the agent respects diverse types of reasoning as valuable and avoids enforcing convergence toward a single “ideal” profile.

While our findings demonstrate the effectiveness of adaptive inquiry-based support, comments from participants who rated the agent as less effective revealed a persistent tension inherent to reflection support systems: the gap between users' expectations for advice and the system's deliberate focus on facilitative inquiry~\cite{reflectionMachines}. 
Some participants ($n$=10 across both conditions) expressed frustration that the agent "is only trained to ask questions" or explicitly stated they expected to "get advice also."

On the one hand, this observation points to a promising direction: systems could offer users flexible control over interaction modes, allowing them to switch between inquiry-based reflection and directive advice as their needs evolve—an approach explored in conversational recommendation systems~\cite{UA3}. Such flexibility could accommodate diverse user preferences and support needs. On the other hand, caution is needed; prioritizing users’ desire for advice should not come at an ethical cost. Prior work suggests that inquiry-driven interactions better support user agency and autonomy, preventing premature advice in high-stakes situations \cite{koralus2025philosophic}. From this perspective, the absence of advice represents a deliberate pedagogical trade-off rather than a system limitation.
Nevertheless, users’ expectations highlight the need for expectation management, suggesting that agents may benefit from explicitly negotiating their role and clearly communicating system boundaries to maintain alignment with system objectives.

Ultimately, in an age where people increasingly rely on LLMs for many tasks, including high-stakes decision making~\cite{altman2025}, there is a risk that these systems may overly influence or bias users. Rather than adapting to the LLM, we aim to contribute to work that fosters and preserves human autotomy without steering decisions or homogenizing reasoning across individuals. 
That said, we emphasize that real-world deployment of such systems requires further research and additional ethical review, particularly to better understand long-term effects and to address the needs of vulnerable users or crisis situations.

\section{Future Work and Limitations}
While the experimental agent tailored its exploitation questions to the user’s answers, the exploration questions were scripted, primarily for experimental purposes. 
This suggests an interesting future direction: exploring how these questions can be personalized to individual users without compromising their ability to facilitate broad exploration. Leveraging human expert input and domain knowledge could guide this adaptation.

A limitation of our evaluation is that LIWC captures linguistic expressions of reflection rather than the underlying sources of those expressions. While LIWC has been widely used as a proxy for psychological and cognitive–affective processes, the markers associated with different processes should not be interpreted as direct evidence of engagement of a specific “system.” 
Accordingly, our analysis focused on shifts in expressed reflective patterns rather than latent states, future work could triangulate these findings with behavioral or physiological measures.

\section{Conclusion}
Grounded in theories of human decision making, we explored whether conversational agents can support an integrative pre-decision reflection process.
Our approach, built directly upon users' observable thought patterns, successfully elicited more integrative reflective language and was perceived as supporting a holistic reflection process, all while respecting the user's reasoning tendencies.
In contrast, the baseline LLM-based agent followed a “one-size-fits-all” approach, pushing all users toward cognition. 
Our novel reflection modeling and adaptation approach, supported by these findings, opens a promising direction for personally meaningful decision support systems, where emotional and intuitive aspects are also treated as complementary forms of reasoning worthy of personalized computational support.

\clearpage
\appendix
\section*{Appendix}



\setlength{\textfloatsep}{8pt plus 2pt minus 2pt}
\setlength{\intextsep}{8pt plus 2pt minus 2pt}
\setlength{\abovecaptionskip}{4pt}
\setlength{\belowcaptionskip}{2pt}

\lstset{
  basicstyle=\ttfamily\footnotesize,  
  frame=single,
  breaklines=true,             
  backgroundcolor=\color{gray!10},
  keywordstyle=\color{blue}\bfseries,
  commentstyle=\color{green!50!black},
  showstringspaces=false,
  aboveskip=6pt,
  belowskip=6pt,
  xleftmargin=3pt,
  xrightmargin=3pt
}

\section{Big Life Decision Topics}
\begin{table}[!htb]
\footnotesize
\centering
\begin{tabular}{@{}ll@{}} 
\toprule
\textbf{Decision Category} & \textbf{Decision Topic} \\ 
\midrule
Career &  Start/Quit a job/position\\
Career &  Start a new business\\ 
Education &  Pursue a degree\\ 
Family & Have/adopt a child \\ 
Family & Care for a family member \\ 
Family & Get pet \\ 
Finances &  Buy home\\ 
Relationships &  Begin/End romantic relationship\\ 
Relationships & Get married/divorced \\ 
Relocation & Move to a new city/country \\ 
\bottomrule
\end{tabular}
\caption{Decision Topics and Categories, fourteen topics in total}
\label{table:topics}
\end{table}

\section{Exploration Questions by Category}
\begin{table}[!htb]
\footnotesize
\centering
\begin{tabular}{@{}lp{0.83\linewidth}@{}}
\toprule
\textbf{Category} & \textbf{Question} \\
\midrule
Internal & What are your gut feelings about \{decision\}? \\
 & When you think about \{decision\}, what does your heart want? \\
 & What emotions come up when you think about making this decision? \\
\cmidrule(lr){1-2}
Experiential & What personal (first-hand) experiences have you had that relate to \{decision\}? \\
 & What stories and experiences from your network (second-hand experiences) can you think of in relation to \{decision\}? \\
 & What lessons or insights from your past experiences might help you in the process of making this decision? \\
\cmidrule(lr){1-2}
External & What external factors are supporting this decision? \\
 & What external constraints are posing challenges in \{decision\}? \\
\bottomrule
\end{tabular}
\caption{Exploration questions used by the experimental agent, grouped by targeted category.}
\label{tab:exploration_questions}
\end{table}


\section{Exploitation Prompt (Experimental Agent)}
\begin{lstlisting}
You are an AI assistant designed to generate follow-up questions that encourage deeper elaboration on a user's argument about {topic}.
Problem:
- The user's argument: {span} lacks depth and requires further exploration.
Your Task:
- Generate a single, concise follow-up question for the argument. 
Question Generation Rules:
- This question can be a **Socratic** question. 
- Ensure that the question is in relation to the decision to be made: {topic} not just a general question.
- Ensure that the question fits within the current context by considering the conversation history. 
- If they are earlier relevant points raised in the conversation history, you should acknowledge them in your question to show that you are aware of the context and ask if the user has more to say.
- Using the conversation history, avoid asking about information the user has already elaborated upon.
- The question should be neutral, non-leading, and avoid introducing new viewpoints.
- The question should be **short**, **clear**, and **directly** reference the original argument: {span}.
- Use a conversational tone
- Ensure the question helps the user **deepen** their reflection, not simply repeat previous information.
Constraints:
- Focus the question on **gathering new insights**, not repeating what has already been said.
- Do not suggest new perspectives or opinions.
- Ask only **one** question per turn.
- Avoid compound or double-barreled questions.
- Stay focused on one idea at a time.
- Keep the question objective and concise.
- Ensure to explicitly include the original text span within the follow-up question.
- The question must fit logically within the flow of the conversation.
\end{lstlisting}

\section{Baseline Agent Instruction}
\begin{lstlisting}
You are an AI assistant designed to support users in reflecting before making a big decision.
Problem:
- The user wants to enhance the **breadth** and **depth** of their reflection by integrating their head, heart, and gut, and by exploring aspects they may have under-considered or overlooked.
Your Task:
- At each turn, generate a single, concise question based on the conversation history to support this process.
Question Generation Rules:
- Ensure that the question is in relation to the decision to be made: {topic} ,not just a general question.
- Ensure that the question fits within the context by considering the conversation history. 
- If they are earlier relevant points raised in the conversation history, you should acknowledge them in your question to show that you are aware of the context
- Using the conversation history, avoid asking about information the user has already said.
- The question should be neutral, non-leading, and avoid introducing new viewpoints.
- The question should be **short**, **clear**.
- Use a conversational tone
- Ensure the question helps the user **deepen** or **broaden** their reflection, not simply repeating previous information.
- This question can be a **Socratic** question.
Constraints:
- Do not suggest new perspectives or opinions.
- Ask only **one** question per turn.
- Avoid compound or double-barreled questions.
- Stay focused on one idea at a time.
- The question must fit logically within the flow of the conversation.
\end{lstlisting}

\section{Post-Questionnaire Items}
\begin{table}[!htb]
\footnotesize
\centering
\begin{tabular}{@{}p{0.72\linewidth}p{0.22\linewidth}@{}}
\toprule
\textbf{Statement} & \textbf{Construct} \\
\midrule
The agent posed questions that helped me integrate head, heart, 
and gut. & Holistic integration \\
\cmidrule(lr){1-2}
The agent posed questions that helped me elaborate further on 
my thoughts. & Elaboration depth \\
\bottomrule
\end{tabular}
\caption{Post-experiment statements used to evaluate perceived support on a 5-point Likert scale (from strongly disagree to strongly agree).}
\label{tab:post_statements}
\end{table}

\newpage

\begin{acks}
We gratefully acknowledge the support of Delft AI Initiative and Designing Intelligence (DI) lab.
This research was (partially) funded by the Hybrid Intelligence Center, a 10-year programme funded by the Dutch Ministry of Education, Culture and Science through the Netherlands Organisation for Scientific Research, \href{https://www.hybrid-intelligence-centre.nl}{Hybrid Intelligence Centre}, grant number 024.004.022.
In addition, it has been partially supported by the BOLD cities initiative.
It was also partially funded by the project RISE: Relapse Intervention and Self-Management (Technology Assisted Self-Management: Preventing Relapse and Crisis by the Severe Mentally Ill Themselves) with file number KICH1.GZ03.21.002 of the research programme KIC - MISSIE Zorg in eigen leefomgeving 2021 which is (partly) financed by the Dutch Research Council (NWO).
\end{acks}

\paragraph{Gen AI Declaration}
Generative AI tools were used to assist with copy-editing, but the authors are fully accountable for the content.

\bibliographystyle{ACM-Reference-Format}
\bibliography{references}

\end{document}